\documentclass[fleqn,10pt,twocolumn]{wlscirep}
\usepackage{epsfig,amsbsy,bm,amssymb}
\usepackage{hyperref}
\usepackage{graphicx}
\usepackage{float}
\usepackage{amsmath}

\DeclareMathOperator*{\SumInt}{%
\mathchoice%
  {\ooalign{$\displaystyle\sum$\cr\hidewidth$\displaystyle\int$\hidewidth\cr}}
  {\ooalign{\raisebox{.14\height}{\scalebox{.7}{$\textstyle\sum$}}\cr\hidewidth$\textstyle\int$\hidewidth\cr}}
  {\ooalign{\raisebox{.2\height}{\scalebox{.6}{$\scriptstyle\sum$}}\cr$\scriptstyle\int$\cr}}
  {\ooalign{\raisebox{.2\height}{\scalebox{.6}{$\scriptstyle\sum$}}\cr$\scriptstyle\int$\cr}}
}

\newcommand{\isum}%
{\mathop{\hbox{$\displaystyle\sum\kern-13.2pt\int\kern1.5pt$}}}
\renewcommand{\r}{{\bm r}}
\renewcommand{\k}{{\bm k}}
\newcommand{\p}{{\bm p}}
  \newcommand{\A}{{\bm A}}
   
  \newcommand{\e}{{\bm e}}
  \newcommand{\ve}{{\bm v}}

\newcommand{\bt}{\begin{tabular}}
\newcommand{\et}{\end{tabular}}

\newcommand{\eref}[1] {(\ref{#1})}
\newcommand{\Eref}[1] {Eq.~(\ref{#1})}

\newcommand{\Fref}[1] {Fig. \ref{#1}}

\newcommand{\br}{\begin{eqnarray*}}
\newcommand{\er}{\end{eqnarray*}}
\newcommand{\ba}{\begin{eqnarray}}
\newcommand{\ea}{\end{eqnarray}}
\newcommand{\be}{\begin{equation}}
\newcommand{\ee}{\end{equation}}

\newcommand{\bp}{\begin{minipage}}
\newcommand{\ep}{\end{minipage}}

\title {Joint probability calculation of the lateral velocity distribution
in strong field ionization process}

\author[1,*]{I. A. Ivanov}
\author[1,2,+]{Kyung Taec Kim}

\affil[1]{Center for Relativistic Laser Science, Institute for
Basic Science, Gwangju 61005, Korea}

\affil[2]{Department of Physics and Photon Science, Gwangju Institute of Science and Technology, Gwangju 61005, Korea}

\affil[*]{igorivanov@ibs.re.kr}
\affil[+]{kyungtaec@gist.ac.kr}

\begin{abstract}
We describe an approach to the description of the time-development of the process of strong 
field ionization of atoms based on the calculation of the joint probability of occurrence of two events, 
event B being finding atom in the ionized state after the end of the laser pulse, 
event A being finding a particular value of a given physical observable at a moment of time inside the laser 
pulse duration.  As an example of such an physical observable we consider lateral velocity component of 
the electron's velocity. Our approach allows us to study time-evolution 
of the lateral velocity distribution for the ionized electron during the interval of the laser pulse duration.
We present results of such a study for the cases of target atomic systems with short range Yukawa and Coulomb 
interactions.
\end{abstract}

\begin{document}

\flushbottom
\maketitle

\thispagestyle{empty}

\section*{Introduction}

The conventional framework for understanding 
the basic features of the tunneling ionization process, known as the Strong Field Approximation (SFA),
was laid out by Keldysh \cite{Keldysh64} and developed and refined
subsequently \cite{Faisal73,Reiss80,ppt,adk1,hhgd,YI,tunr,tunr2}. In this framework the
tunneling regime of ionization is defined as the regime characterized by the small values of the so-called 
Keldysh parameter $\gamma=\omega\sqrt{2|\varepsilon_0|}/E$ (here $\omega$, $E$ and $|\varepsilon_0|$ are the frequency, field strength and ionization potential
of the target system expressed in atomic units). 

A number of nonlinear phenomena, such as  above-threshold ionization 
(ATI), high harmonic generation  (HHG) and non-sequential
double ionization (NSDI) may occur in this regime. The most essential features of these 
processes can often be understood with the help of the well-known simple man model (SMM) \cite{hhgd,Co94,kri,tipis,arbm}, which represents electron's motion after the ionization event occurs 
as entirely classical. This fact underlies a variety 
of highly efficient semi-classical approaches \cite{2step,xv6,tipis,cusp3,arbm,tipis_naft,landsman2015,qorb1}, 
in which the 
electron ionization event 
is treated quantum-mechanically, with electron's ionization occurring near the local peaks of the
electric field, and the 
subsequent electron motion treated classically or semi-classically. These approaches
include the well-known TIPIS model \cite{tipis,cusp3,cmtc1}, 
the quantum trajectory Monte Carlo model (QTMC) \cite{qorb1}, semi-classical 
two-step model \cite{Shvetsov-Shilovski2016} or 
Coulomb quantum orbit strong-field
approximation (CQSFA) \cite{xv7,xv8}.  

Despite these semi-classical methods giving often impressive agreement with more 
rigorous calculations based on the solution of the time-dependent Schr\"odinger equation
(TDSE), it is not quite clear what the very notion of the 'ionization event' means from
the point of view of the conventional quantum mechanics (QM). Indeed, this notion
is not easy to define even in classical physics. This problem manifests itself in 
the treatment of ionization in the standard QM framework in the following way. QM provides us
with only a wave-function to describe evolution of the system in the laser field, and 
for the moments of time inside the interval of the laser pulse duration, when 
the wave-packet describing ionized electron has not left the atom yet,
it is difficult to unambiguously single out the part of the wave-function describing ionized
electron from the total wave-function of the system. 
At first glance such a separation of the total TDSE wave-function 
into the bound component and the ionized wave-packet arises quite naturally 
in the SFA or the Perelomov-Popov-Terent'ev (PPT) 
approaches \cite{Keldysh64,Faisal73,Reiss80,ppt,tunr,adk}. We should note,
however, that this splitting of the total wave-function in two components in these theories,
although extremely useful and physically appealing, is not quite rigorous. That can been seen,
for instance, by noting that this splitting is not gauge-invariant.  It is this fact
that is ultimately responsible for the lack of the gauge-invariance of the SFA or PPT approaches \cite{tunr2}.
To see more clearly the origin of the problem we will recapitulate briefly, following the 
work \cite{sfam}, the theoretical foundations of 
both approaches, which can done most easily by using the 
notion of the unitary time-evolution  operator $\hat U(t,\tau)$ which maps  
state of the system  at time $\tau$ to the state at time $t$.
For a system with the time-dependent Hamiltonian operator $\hat H(t)=\hat H_A(t) + \hat B(t)$.
the integral Dyson equation for the evolution operator $\hat U(t,\tau)$ 
can be written in two equivalent forms as:

\begin{eqnarray}
\hat U(t,0)&=& \hat U_A(t,0) -i\int\limits_0^t \hat U_A(t,\tau) \hat B(\tau)\hat U(\tau,0)\ d\tau \nonumber \\
\hat U(t,0)&=& \hat U_A(t,0) -i\int\limits_0^t \hat U(t,\tau) \hat B(\tau)\hat U_A(\tau,0)\ d\tau \,
\label{p1b} 
\end{eqnarray}

where $ \hat U_A(t,\tau)$ is the time-evolution operator describing evolution driven by the Hamiltonian 
operator $\hat H_A$. In the case of the Hamiltonian  which describes atomic or molecular system interacting with
the electromagnetic field: $\displaystyle \hat H(t)=\hat T + \hat V + \hat H_{\rm int}(t)$,
where $\hat T$ is the kinetic energy operator, $\hat V$- the potential energy operator, and
$\hat H_{\rm int}(t)$ describes the interaction of the system and the electromagnetic field, 
two useful partitions of the total Hamiltonian $\hat H(t)$ are:

\begin{eqnarray}
\hat H(t)&=& \hat H_{\rm atom} + \hat H_{\rm int}(t) \nonumber \\
\hat H(t)&=& \hat H_{F}(t) + \hat V \ ,
\label{partb}
\end{eqnarray}

where $\hat H_{\rm atom}= \hat T + \hat V$ is the field-free Hamiltonian of the system,
$\hat H_F(t)=  \hat T + \hat H_{\rm int}(t)$ is the so-called Volkov Hamiltonian.
Using these two partitions in the Dyson equations \eref{p1b} , one obtains two
equations:

\begin{eqnarray}
\hat U(t,0)&=& \hat U_0(t,0) -i\int\limits_0^t \hat U(t,\tau) \hat H_{\rm int}(\tau)\hat U_0(\tau,0)\ d\tau \nonumber \\
\hat U(t,0)&=& \hat U_F(t,0) -i\int\limits_0^t \hat U_F(t,\tau) \hat V \hat U(\tau,0)\ d\tau \ , 
\label{p2b} 
\end{eqnarray}

where $\displaystyle  \hat U_0(t,\tau) = \exp{\left\{-i\hat H_{\rm atom}(t-\tau)\right\}}$ is the 
evolution operator describing evolution driven by the time independent field-free Hamiltonian 
$\hat H_{\rm atom}$ and $\hat U_F(t,\tau)$ is the so-called Volkov evolution operator with 
a simple known analytical form \cite{tunr2} which
describes quantum evolution driven by the Volkov Hamiltonian $\hat H_F(t)$. 

Equations \eref{p2b} are exact but are not very useful, being just as difficult to solve as  
the original TDSE with which we started. The widely used and important SFA and PPT 
approximations are obtained if we substitute  $\hat U_F(t,\tau)$ for $\hat U(t,\tau)$ under the integral in the first 
Dyson equation \eref{p1b} and 
$\hat U_0(\tau,0)$ for $\hat U(\tau,0)$ in the second equation \eref{p1b}:

\begin{eqnarray}
\hat U_{SFA}(t,0)&=& \hat U_0(t,0) -i\int\limits_0^t \hat U_F(t,\tau) \hat H_{\rm int}(\tau)\hat U_0(\tau,0)\ d\tau \nonumber \\
\hat U_{PPT}(t,0)&=& \hat U_F(t,0) -i\int\limits_0^t \hat U_F(t,\tau) \hat V \hat U_0(\tau,0)\ d\tau \ . 
\label{p22b}
\end{eqnarray}

If we use the expression for the evolution operator in the first equation \eref{p22b} 
to find the wave-function at the moment of time $t=T_1$ at the end of the laser pulse for a system which was initially in the 
field free state $\phi_0$, we obtain an approximate wave-function: $\Psi_{SFA}(T_1)= \hat U_{SFA}(T_1,0)\phi_0$.
Projecting this wave-function on a plane-wave state $|\k\rangle $ and dropping the term $\langle \k|\hat U_0(T_1,0)|\phi_0\rangle$,
one obtains the well-known expression for the ionization amplitude used by Keldysh \cite{Keldysh64}, provided we 
use the length form for the interaction Hamiltonian $\hat H_{\rm int}(t)$. If, instead, we use
the velocity gauge to describe field-atom interaction, we obtain the expression for the ionization amplitude used in the 
well-known Strong Field Approximation (SFA) theory \cite{Faisal73,Reiss80}. Using, on the other hand,   
the approximate evolution operator in the second equation \eref{p22b} to evaluate the 
wave-function at the moment of time $t=T_1$ at the end of the laser pulse for a system which was initially in the 
field free state $\phi_0$, we
obtain an approximate wave-function $\Psi_{PPT}(T_1)= \hat U_{PPT}(T_1,0)\phi_0$. Projecting this wave-function
on a plane-wave state $|\k\rangle $ and dropping the term $\langle \k|\hat U_F(T_1,0)|\phi_0\rangle$ which does not contribute
to the probability current, one obtains the expression for the ionization amplitude used in the 
Perelomov-Popov-Terentiev (PPT) theory \cite{ppt}. 

We see, thus, that both in the PPT and SFA approaches we obtain expressions for the ionization
amplitudes by omitting certain terms from the total wave-function. These terms, however, are 
different in the two approaches and, moreover,  it is this omitting that breaks the gauge invariance which is
of course present in the complete theory. 
Under a gauge transformation only the total wave-function of a system is transformed in a way ensuring 
gauge invariance of the final results for the observables. The parts of the wave-function 
obtained by splitting it into different components will not, in general,  possess this property. 
One needs some, therefore, additional theoretical ingredients in the theory, which may allow to define the notion
of the ionized part of the wave-function. An example of such a theoretical development is
provided by the well-known back-propagation technique \cite{Nie,Nie1,xv1,xv2,xv5}. 
This procedure allows to single out only ionized electron and to follow 
its prehistory back in time by constructing the ionized wave-packet from the part of the coordinate 
wave-function localized  
far from the nucleus for times long after the end of the driving pulse.

In the present work we describe another procedure allowing to achieve such a separation of the
total wave-function describing the evolution of an atom in the laser field into the ionized and 
non-ionized components. We will use the notion of the conditional and joint probabilities for 
this purpose. 
In the Section Theory below we describe our theoretical procedure. We will illustrate this 
procedure by applying it to a study of the evolution of the lateral velocity distribution (i.e. distribution 
of the velocity components in the directions orthogonal to the polarization vector of the 
driving field) of ionized electrons for times inside the laser pulse duration.
These applications are described in the Section Results below. We conclude in the 
Section Conclusion making a brief summary of the results and future perspectives.
Atomic units (a.u.) are used throughout the paper.

\section*{Theory}

\subsection*{Joint probability in Quantum Mechanics}

We begin by presenting a short summary of the theory of the conditional and joint probabilities in
QM following work \cite{cond1}. Suppose we have a system described
by the state vector $|\Psi\rangle$, and two observables 
with the corresponding Hermitian operators $\hat A$ and $\hat B$. The probability to
find the system upon measurement in an eigenstate of $\hat B$  belonging to a
spectral interval $\Delta_B$ of the operator $\hat B$, can be computed, as is well known, as 
follows:

\be
P(\Delta_B)= \langle \Psi|\hat Q(\Delta_B)|\Psi\rangle  \ ,
\label{c1}
\ee 

where $\hat Q(\Delta_B)$ is the Hermitian projection operator which can be expressed in terms of 
the eigenstates $|\lambda\rangle$ of the operator $\hat B$ as:  
$\displaystyle \hat Q(\Delta_B)= \SumInt\limits_{\Delta_B} 
|\lambda\rangle\langle\lambda|\ d\lambda$. Analogous formula, 
where we will have to use a projection operator $\hat Q(\Delta_A)$, which can be
similarly expressed in terms of the eigenstates of the operator $\hat A$, can
be written, of course, for the probability $P(\Delta_A)$ 
to find the system in an eigenstate of $\hat A$  belonging to a
spectral interval $\Delta_A$ of the operator $\hat A$.

Suppose now that we are interested in the joint probability $P(\Delta_A \& \Delta_B)$ of finding 
observable $A$ in the spectral interval $\Delta_A$, and observable $B$ in the spectral interval $\Delta_B$.
One might try to define such a joint probability by an expression:

\be
P(\Delta_A \& \Delta_B) = \langle \Psi|\hat Q(\Delta_A)\hat Q(\Delta_B)|\Psi\rangle  \ .
\label{c2}
\ee 

This definition would make perfect sense if the quantum projection 
operators $\hat Q(\Delta_A)$ and $\hat Q(\Delta_B)$ 
were classical quantities or at least quantum commuting operators. Unfortunately, this
is not necessarily always the case. If the operators $\hat A$ and $\hat B$ do not 
commute, the corresponding projection operators $\hat Q(\Delta_A)$ and $\hat Q(\Delta_B)$
do not commute either, the operator product $\hat Q(\Delta_A)\hat Q(\Delta_B)$ in \Eref{c2} 
becomes non-Hermitian and the joint probability defined by \Eref{c2} is generally complex-valued. 
This is, of course, another manifestation of the well-known difficulty that one encounters
when trying to assign a meaning to the joint probability distributions of the 
observables described by non-commuting operators using the quasi-probability 
distributions, such as Wigner and Husimi distributions \cite{qp0,qp1,xv4,qp2} which 
may not necessarily be strictly positive. With \Eref{c2} we have the same problem in 
another disguise, it can give, as we see, complex-valued joint probability distributions.
It can be argued \cite{cond1,cond2,cond3} that such quasi-probability distributions, 
which are not strictly positive, can be incorporated in the
framework of the QM, and that they appear naturally in the description of the classically
forbidden processes such as the tunneling process \cite{cond2}. Negative probabilities
are, however, difficult to interpret, and many physicists are somewhat reluctant in 
accepting such a notion. 

It is the tunneling 
process due to atomic ionization in the external electric field that interests us in the present work. 
We will see below that in some circumstances \Eref{c2} gives 
us joint probability distributions which are
perfectly legitimate in an ordinary
probabilistic sense, and which may 
provide useful information about the 
development of the tunneling ionization process.

\subsection*{Joint probabilities for the strong field ionization process}

We will be studying below ionization of a single-electron atom in a strong 
electromagnetic field. The evolution of the system in time is described by
the three-dimensional TDSE:

\begin{equation}
i {\partial \Psi(\r,t) \over \partial t}=
\left(\hat H_{\rm atom} + \hat H_{\rm int}(t)\right) \Psi(\r,t) \ ,
\label{tdse}
\end{equation}

with $\displaystyle H_{\rm atom}= {\hat\p^2\over 2}+V(r)$- field free atomic Hamiltonian,
and $\hat H_{\rm int}(t)$- interaction Hamiltonian describing atom-field interaction. We
use the length gauge for the latter operator:

\be
\hat H_{\rm int}(\r,t) = \r\cdot{\bm E}(t) \,
\label{hint} 
\ee

where ${\bm E}(t)$ is electric field of the pulse. We assume that pulse is linearly polarized along 
$z$-axis. The pulse is defined
in terms of the vector potential: $\displaystyle \bm{E}(t)=-{\partial \A(t)\over \partial t}$,
where:

\be
\A(t)= -{\hat \e_z}{E_0\over\omega}\sin^2{\left({\pi t\over T_1}\right)}\sin{\omega t} \ ,
\label{vp}
\ee

where $T_1$ is the total pulse duration, which we choose to be one optical cycle (o.c.): $T_1=2\pi/\omega$,
for the majority of the calculations reported below. We use a single cycle pulse for purely computational
reasons as the calculations become rather time-consuming for longer pulses, 
our approach can be generally applied for long pulses as well.
We will present below results for different
values of the pulse base frequency $\omega$, peak field strength $E_0$ and different atomic potentials 
$V(r)$. 

To apply \Eref{c2} and the notion of the joint probability distribution in QM to the process 
of the tunneling ionization we should first specify the operators $\hat A$ and $\hat B$ in 
this equation. We note first that it is by no means necessary that both projective 
measurements in \Eref{c2} are performed at the same moment of time. We may well assume that
one of the projective measurements (let's say the measurement described by the projection 
operator $\hat Q(\Delta_B)$) is performed at the time $t_2$, while the measurement 
described by the operator $\hat Q(\Delta_A)$ is performed at the time $t_1$, with 
$t_2>t_1$. To modify \Eref{c2} accordingly it is convenient to use the
Heisenberg picture of the QM, in which the state vector $\Psi=\phi_0$ is 
independent of time (here $\phi_0$ is the initial state of the system), while the projection 
operators $\hat Q(\Delta_A)$ and $\hat Q(\Delta_B)$ evolve in time.
With the use of the Heisenberg picture the generalization of the \Eref{c2} for the case 
of different times is straightforward:

\be
P(\Delta_A(t_1) \& \Delta_B(t_2)) = 
\langle \phi_0|\hat Q^H(\Delta_A,t_1)\hat Q^H(\Delta_B,t_2)|\phi_0\rangle  \ ,
\label{c3}
\ee 

with

\begin{eqnarray}
\hat Q^H(\Delta_A,t_1)&=& \hat U(0,t_1) \hat Q(\Delta_A)\hat U(t_1,0) \nonumber \\
\hat Q^H(\Delta_B,t_2)&=& \hat U(0,t_2) \hat Q(\Delta_B)\hat U(t_2,0) \ ,
\label{hp}
\end{eqnarray}

where $\hat U(t,0)$ is the evolution operator, driving quantum evolution of the system,
and $|\phi_0\rangle $ is the initial atomic state (which we assume to be the ground state of the 
field-free atomic Hamiltonian $ H_{\rm atom}$.

Definition \Eref{c3} of the joint probability distribution is quite similar to the definition of the 
two-time correlation function \cite{autoh2,autoh1,auto1,auto3}, describing correlations between two 
observables with corresponding quantum mechanical operators $\hat Q(\Delta_A)$ and $\hat Q(\Delta_B)$
at times $t_1$ and $t_2$, respectively. Correlation functions can be employed
\cite{corn1,corn2} to study correlations between different observables in the strong field ionization
process. Such a study may provide useful information about presence or absence of the correlations between
observables at different moments of time. On the other hand, the two-time correlation functions,
being generally complex-valued objects, do not have direct physical interpretation. The approach we describe in 
the present work differs from the approach based on the two-time correlation functions analysis 
in one important aspect. We assume that the quantum operators  
$\hat Q^H(\Delta_A,t_1)$ and $\hat Q^H(\Delta_B,t_2)$ in \Eref{c3} are  
are not arbitrary operators, but quantum mechanical projection operators. This implies that
both $\hat Q^H(\Delta_A,t_1)$ and $\hat Q^H(\Delta_B,t_2)$  are 
positive-definite Hermitian operators. Moreover, if
$\hat Q^H(\Delta_A,t_1)$ and $\hat Q^H(\Delta_B,t_2)$ are commuting operators, the operator product 
$\hat P=\hat Q^H(\Delta_A,t_1)\hat Q^H(\Delta_B,t_2)$ in \Eref{c3} is again a positive definite 
Hermitian operator with positive real expectation value. It is easy to see,
in fact, that $\hat P$ is then a projection operator, satisfying $\hat P^2=\hat P$.
If $\hat Q^H(\Delta_A,t_1)$ and $\hat Q^H(\Delta_B,t_2)$ commute we can, therefore, assign direct 
physical meaning of the joint probability to the quantity defined by the \Eref{c3}.
We will now specify the observables $A$ and $B$ and the projection operators 
$\hat Q(\Delta_A)$ and $\hat Q(\Delta_B)$ in \Eref{hp}.

\subsubsection*{Choice of the observable $B$}

We will assume from now on that
in \Eref{c3} the later moment of time $t_2=T_1$, the moment of time when the laser pulse \eref{vp}
is switched off. For the projection operator $\hat Q(\Delta_B)$ we will use the projection operator 
projecting the state vector on the continuous part of the energy spectrum of the atomic Hamiltonian 
$ H_{\rm atom}$,and by the spectral interval $\Delta_B$ we will understand the whole range of 
positive electron energies of the continuous spectrum of the field-free atomic Hamiltonian.
In other words, $Q(\Delta_B) =\hat Q_{c}$, where $\hat Q_c$ 
is the projection operator on the continuous spectrum of the
field-free atomic Hamiltonian, so that for any state vector $|\Psi\rangle$:

\be
\hat Q_c |\Psi\rangle = |\Psi\rangle- 
\sum\limits_{bound\atop states} \langle\phi_b|\Psi\rangle |\phi_b\rangle \ ,
\label{q}
\ee

where the sum on the r.h.s includes all bound states of the field-free atomic Hamiltonian
$ H_{\rm atom}$. It is worth discussing briefly what we have achieved by choosing this particular
form of the projection operator $\hat Q(\Delta_B)$. Observable $B$ with this choice of the 
projection operator provides basically the answer to the question: will the electron be ionized or not,
it might be pictured as an observable having the value one if electron is ionized, and zero otherwise. 
The joint probability distribution \eref{c3} (provided it gives a legitimate probability distribution,
we will touch on this question later) gives us therefore a joint distribution of the 
observable $A$ at the time $t_1$ inside the laser pulse duration upon the condition that electron
will be ionized in the future. By using the joint probability approach and our choice of the 
observable $B$ we are able, therefore, to separate ionized and non-ionized electrons. 
As we noted in the Introduction such a separation is by no means trivial and
requires some additional theoretical notions, such as 
the notions based on different ionization criteria employed in the 
back-propagation technique \cite{Nie,Nie1,xv1,xv2,xv5}. 
We can look at the \Eref{c3} as yet another way to achieve this separation and
study prehistory of the ionized electron based on the notion of joint probability distribution 
and the choice of the observable $B$ we made above. 

\subsubsection*{Choice of the observable $A$}

As far as the observable $A$ in \Eref{c3} is concerned, we can choose, 
in principle, any observable which will provide useful information about the ionization
process. Our only limitation is that the choice of $A$ must be such that the joint 
distribution \eref{c3} be at least approximately legitimate probability distribution.
What it means in practice is that we must choose $A$ so that the imaginary part
of the distribution \eref{c3} be small comparing to its real part for all values of 
$A$ of interest, i.e. for the bulk of the distribution where the joint probability has 
non-negligible values. If this condition is satisfied,
the projection operators in \Eref{c3} will (approximately) commute, and 
\Eref{c3} will automatically give a joint distribution which will be positive at 
least for the bulk of the distribution. We will, of course,  still have non-physical 
negative probabilities on the 
edges of the distribution, where the joint probability is small. That is inevitable unless
the projection operators $\hat Q(\Delta_A,t_1)$ and $\hat Q(\Delta_B,t_2)$ in \Eref{c3} strictly 
commute,  but that is presumably not very important. 

That  was the strategy we pursued in choosing the observable $A$. It turns out that, while
it is not generally possible to choose a physically interesting observable 
$A$ so that imaginary part of the joint distribution
\eref{c3} be always small, it is possible to achieve this goal for at least some values of 
time $t_1$ inside the pulse. That can still provide us with valuable information about 
development of the ionization process and evolution of the observables characterizing ionized 
electrons inside the interval of the pulse duration .

An observable $A$ which we will study below in detail is the lateral component of electron's
velocity, i.e component of the velocity perpendicular to the polarization vector 
of the laser pulse ($z-$direction for the geometry we use in \Eref{vp}). 
It will be convenient in the following to use a cylindrical coordinate system 
$(\rho,\phi,z)$ in the vector space of electron's velocities $\ve$
with the  longitudinal axis along the $z-$ direction. 
Let us consider a set of regions $\Omega_k$ of the electron's velocity space, 
where integer $k=0,1\ldots$,
such that the components of the vector $\ve$ in the cylindrical coordinate system we 
introduced above satisfy for any integer $k$: 
$\displaystyle \Omega_k= ( kd \le v_\rho < (k+1)d; -\infty < v_z < +\infty)$.
In other words, $\Omega_k$ is the space between two cylinders of infinite height and 
radii of $kd$ and $(k+1)d$. In the calculations below we use $d=0.1$ a.u. 
Parameter $d$ affects the overall 'resolution' in the velocity 
space which we may hope to achieve in the framework of our approach. We should choose is so 
as to not to loose any important fine details of the lateral velocity distributions. 
We will justify the choice of the value $d=0.1$ a.u. that we made below. 
Let $\chi_{\Omega_k}(\ve)$ be the characteristic function of 
the region $\Omega_k$ (i.e. $\chi_{\Omega_k}(\ve)=1$ for $\ve$ inside $\Omega_k$, 
$\chi_{\Omega_k}(\ve)=0$ otherwise). We define now the action of the 
projection operator $\hat Q_k$ on a state vector $|\Psi\rangle$ as follows. 
We first perform Fourier transform of the coordinate wave-function $\Psi(\r)$ obtaining
momentum space wave-function $\tilde \Psi(\ve)$. We define now 
$\hat Q_k|\Psi\rangle $ as an 
inverse Fourier transform of $\tilde \Psi(\ve)\chi_{\Omega_k}(\ve)$. With this definition
$\langle\Psi|\hat Q_k|\Psi\rangle $ is clearly the probability to detect electron's velocity
in the region $\Omega_k$ of the velocity space we defined above. The regions 
$\Omega_k$ with $k=0,1\ldots $ in the velocity space cover all the space and do not intersect, 
they satisfy, therefore, $\hat Q_i\hat Q_j=\delta^i_j \hat Q_i$. 
We now identify $\hat Q_k$ with $\hat Q(\Delta_A)$
where, for every integer $k$ it is  understood that spectral region $\Delta_A$ coincides with 
the region $\Omega_k$ in the velocity space. In physical terms, distribution 
\eref{c3} in which we substitute projection operator $\hat Q_k$ for observable $A$  
gives us a probability of finding electron's velocity anywhere in the region 
$\Omega_k$ of the velocity space at the moment of time $t_1$ provided  that
electron is found to be ionized at the end of the laser pulse. 

\subsection*{Calculation of the joint probability}

Following the discussion presented in the two preceding subsections, we 
rewrite the formula \eref{c3} for the joint probability using more
compact notation:

\begin{eqnarray}
P(\Omega_i(t) \& Q_c(T_1)) & = &  
\langle \phi_0|\hat Q_i^H(t)\hat Q_c^H(T_1)|\phi_0\rangle  \nonumber \\
& = & \langle \Psi(t)|\hat Q_i \hat U(t,T_1)\hat Q_c|\Psi(T_1)\rangle   \ ,
\label{c4}
\end{eqnarray}

where in the first line we used the Heisenberg representation for the projection 
operators (as in \Eref{c3}), while to derive the second line we used the transformation
equations \eref{hp} and the property $|\Psi(t)\rangle = \hat U(t,0)|\phi_0\rangle$ of the 
evolution operator, so that $ |\Psi(t)\rangle $ is the solution of the TDSE obtained at the
moment $t$ from the initial (ground) atomic state $|\phi_0\rangle$. It is the 
quantity on the r.h.s of the second line of \Eref{c4} that 
was actually computed in our calculations. We did it as follows. 
The TDSE \eref{tdse} is propagated forward in time
on the interval of the laser pulse duration $(0,T_1)$ 
starting with an initial atomic state $|\phi_0\rangle$,
thus obtaining the state vector  $|\Psi(T_1)\rangle$ describing atomic system 
at the end of the laser pulse.
Next, we apply the operator $\hat Q_c$ to the vector $|\Psi(T_1)\rangle$ and 
propagate back in time both the resulting vector and the original
vector $|\Psi(T_1)\rangle$, obtaining two time-dependent state-vectors:
$|\Psi_1(t)\rangle= \hat U(t,T_1)\hat Q_c|\Psi(T_1)\rangle$ and 
$|\Psi(t)\rangle= \hat U(t,T_1)|\Psi(T_1)\rangle$.
For any given moment $t$ inside the interval of the pulse duration we can now 
compute the joint probability defined by the \Eref{c4} as a matrix element
$\langle \hat Q_i \Psi(t)|\Psi_1(t)\rangle$ (we used here Hermicity of 
the operator $\hat Q_i$). 
This calculation requires multiple solutions of the 
TDSE for different regions $\Omega_k$ defining the projection operators $\hat Q_k$.
The 3D TDSE was solved numerically using the procedure we tested and described
in detail in \cite{cuspm,circ6,ndim}. The procedure relies on representing
the coordinate wave-function as a series of spherical harmonics with 
quantization axis along the pulse polarization direction.
Spherical harmonics with orders up to $L_{\rm max}=70$ were used.
The radial variable is treated by discretizing the TDSE on a grid with the step-size
$\delta r=0.05$ a.u. in a box of the size $R_{\rm max}=200$ a.u. Necessary checks were 
performed to ensure that for these values of the parameters $L_{\rm max}$ and $R_{\rm max}$
convergence of the calculations has been achieved. The solution of the 3D TDSE 
was propagated both forward and backward in time using the matrix
iteration method \cite{velocity1}.
We did calculations for two atomic systems: system with the field-free dynamics governed by 
the short range (SR) Yukawa potential $\displaystyle V(r)=-{1.9083e^{-r}\over r}$ and the hydrogen atom with 
Coulomb potential $\displaystyle V(r)=-{1\over r}$. Both these systems have the same ionization
potential $I_p=0.5$ a.u. and we use their ground $s$-states as initial states $|\phi_0\rangle$ 
in the calculations below.

The joint probability distribution in \Eref{c4} depends on the time 
moment $t$ and the region $\Omega_k$ in the velocity space. We remind that 
this region was defined above as the volume in the velocity space between two
cylinders of radii $kd$ and $(k+1)d$ where $k=0,1,\ldots$:
$\displaystyle \Omega_k= ( kd \le v_\rho < (k+1)d; -\infty < v_z < +\infty)$,
and we used $d=0.1$ a.u.

By introducing a variable $v_{\perp}= (k+1/2)d$ 
the joint probability distribution in \Eref{c4} can be considered as a (discretized)
function of the lateral velocity $v_{\perp}$ and time $t$ computed on a lateral 
velocity grid  
$v_{\perp}= (k+1/2)d$, $k=0,1,\ldots$. We will use this fact to simplify the notation yet a bit more.
We define now a function:

\be
G(v_{\perp},t)= {\Im( P(\Omega_k(t) \& Q_c(T_1)))^2 \over |P(\Omega_k(t) \& Q_c(T_1))|^2} \ ,
\label{ver}
\ee

which tells whether the imaginary part of the joint probability can be considered as 
small for a given moment of time $t$. The choice of the particular form of the function 
$G(v_{\perp},t)$ is not unique, of course, and even is not very important.  
We only need a convenient measure allowing us
to gauge the relative magnitudes of the real and imaginary parts of the joint probability defined 
in \Eref{c4} and to judge to what extent the 
joint probability is a real quantity. The function defined in \Eref{ver} is just the 
simplest choice allowing to achieve this purpose. 

As we discussed above, in general, we cannot expect that 
for a given $t$ $G(v_{\perp},t)$ vanishes for all $v_{\perp}$ altogether. We can, however, have
$G(v_{\perp},t)$ small for the bulk of the $v_{\perp}$-distribution, i.e., in the region
of lateral velocities where \Eref{c4} has non-negligible values. In such situations
we can use \Eref{c4} to define a physically legitimate probability distribution
for lateral velocities $v_{\perp}$. Our strategy, therefore, is to look at the 
regions in the $(t,v_{\perp})$-plane where we have $G(v_{\perp},t)\ll 1$. If 
such regions in the $(t,v_{\perp})$-plane can be found for time $t$ inside the laser pulse
duration we will obtain a means of calculating a physically sensible 
probability distribution for lateral velocities $v_{\perp}$.
We can compare the distributions thus obtained with 
the theoretical predictions obtained in the framework of the SFA.
The well-known SFA expression for the lateral velocity distribution reads \cite{tunr2,tunr}:

\be 
P^{SFA}(v_{\perp})= A \exp\left\{-{(2I)^{1\over 2}v_{\perp}^2\over E_0}\right\} \ ,
\label{sfa}
\ee

where $I$ is the ionization potential, $E_0$- electric field strength,
and $v_{\perp}^2= v_x^2+ v_y^2$ for the geometry
we employ. We do not specify the constant $A$ in this expression as we will be interested
below in the distribution shape which is described by the exponential function. 
Expression \eref{sfa}  
gives us the final velocity distribution at the detector when the 
laser pulse is gone,
but it can also be used as a plausible expression for the lateral velocity distribution 
within interval of the pulse duration, as is done 
e.g., in the semi-classical simulations \cite{tipis,cusp3,cmtc1,Shvetsov-Shilovski2016}. 
The rational behind that is that a linearly polarized 
electric field does not affect the distribution in the lateral direction during the electron's 
motion subsequent to the ionization event. The width of the lateral velocity SFA distribution \eref{sfa}
can, therefore, be used as a guide in choosing the value for the parameter $d$ we used to foliate the velocity space.
For the field strengths of the order of $E_0\approx 0.07$ we are interested in, the full width at half maximum 
(FWHM) of the distribution \eref{sfa} is approximately $0.45$ a.u., so our choice of $d=0.1$ a.u. allows us
to foliate the velocity space into the layers thin enough so as not to loose important details of the 
velocity distribution. The accuracy could be increased by using a smaller value of the 
parameter $d$ at the expense of additional computing time, but 
as we shall see below, with the foliation parameter $d=0.1$ a.u. the lateral
velocity distribution obtained in the framework of our approach agrees pretty well with the 
SFA expression \eref{sfa} for the short-range interaction, confirming the overall consistence and
accuracy of the procedure. 

To be able to compare the SFA distribution 
\eref{sfa} to the joint distribution \eref{c4} we should take into account the fact that
\eref{c4} refers to the joint probability of finding electron's velocity in the 
region $\Omega_k$: $\displaystyle \Omega_k= ( kd \le v_\rho < (k+1)d; -\infty < v_z < +\infty)$
between the two cylindrical surfaces, while expression \Eref{sfa} refers to the Cartesian 
volume element in the velocity space. To take into account this difference of the 
volume elements we can observe that the volume $\Omega_k$ is proportional to $v_{k,\perp}$,
where $v_{k,\perp}= (k+1/2)d$. We must therefore divide the joint probability in
\Eref{c4} by $v_{k,\perp}$ obtaining the distribution:

\be
P(v_{\perp},t)= {P(\Omega_k(t) \& Q_c(T_1)) \over v_{k,\perp} } \ ,
\label{vdist}
\ee

which refers to the Cartesian volume element in the velocity
space and can, therefore, be compared with the results predicted by the SFA formula \eref{sfa}. We will perform
such a comparison below, by fitting the results we will obtain 
using \Eref{vdist} with analytical formulas. We will
employ two types of fit, one for the SR Yukawa 
potential and another for Coulomb potential. The SR 
interactions fall into the domain of the validity of 
the standard SFA \cite{tunr2,tunr}. For the SR interaction,
therefore, we will fit our results using the fitting expression
based on the SFA formula \eref{sfa}:

\be 
P^{SR}(v_{\perp})= A \exp\left\{-{v_{\perp}^2\over\beta}\right\} \ ,
\label{fitsr}
\ee

where $A$ and $\beta$ are used as fitting parameters.
Parametrization in terms of the parameter 
$\beta$ is convenient as for both Yukawa and Coulomb
systems we have $I=0.5$ a.u. Therefore, for the 
lateral velocity distributions having shapes similar to 
the ones predicted by the SFA formula \eref{sfa}, parameter
$\beta$ should be approximately equal to the peak field 
strength $E_0$ of the laser pulse.

For the Coulomb potential situation is a 
bit more complex. Due to the well-known effect
of the Coulomb focusing \cite{cusp2}, lateral velocity
distribution acquires a cusp at zero transverse velocity.
Such a cusp-like structure can be described
using an expression \cite{cuspm,cuspm1,cusp3}. 

\be 
P^{C}(v_{\perp})= A \exp\left\{-{|v_{\perp}|^{\alpha}\over\beta}\right\} \ ,
\label{fitc}
\ee
where $A$, $\alpha$ and $\beta$ are used as fitting parameters.
Depending on the value of the non-integer parameter 
$\alpha$ the distribution function in \eref{fitc} exhibits at
$v_{\perp}=0$ a discontinuity in first or higher order derivatives, thus serving as a
model of the cusp-like behavior.

The lateral velocity distributions that we compute using
\Eref{vdist} pertain to the moments of time insider the
interval of the pulse duration. We will compare them
also to the asymptotic (obtained in the limit of large times) velocity distributions $P^{as}(v_{\perp})$,
which we compute using the standard prescription, by 
projecting the wave-function at the end of the pulse
on the scattering states of the field-free atomic 
Hamiltonian (with ingoing boundary conditions) :

\be
P^{as}(v_{\perp}) = 
\int |\langle|\phi^-_\ve|\Psi(T_1)\rangle|^2\ dv_z \ .
\label{as}
\ee

Such a comparison is of interest, especially for the
Coulomb potential, as it shows how cusp in the lateral
velocity distribution develops in time. It is known 
that the Coulomb atomic system needs to evolve for some time
for the cusp to develop \cite{cuspo}. Comparing 
asymptotic lateral velocity distribution 
\eref{as} to the distributions \eref{vdist} we will
be able to have a glimpse at how this development of the
cusp actually occurs.

\section*{Results and discussion}

\subsection*{Yukawa potential}

\begin{figure*}[!]
\resizebox{150mm}{!}{\epsffile{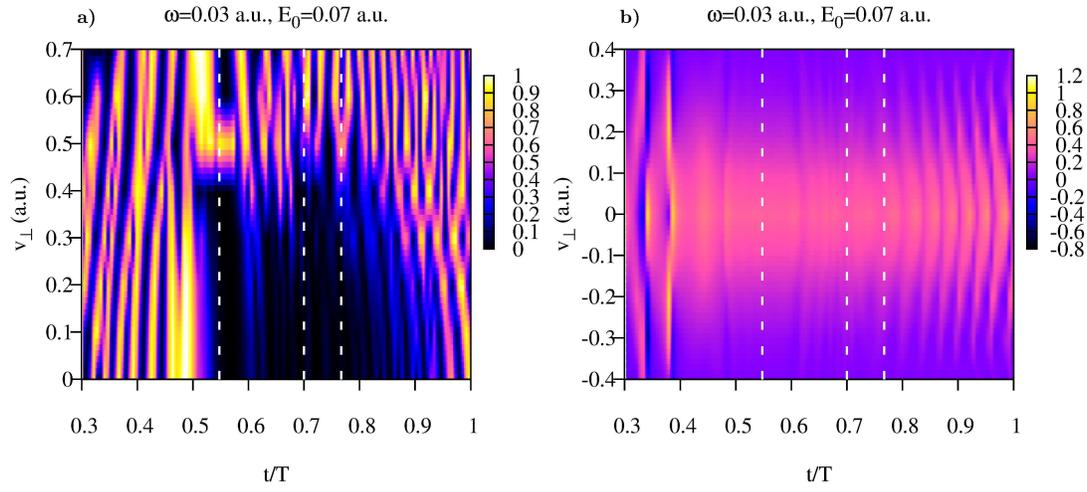}} 
\caption{(Color online) Yukawa potential. (a): Function $G(v_{\perp},t)$ 
(b): Distribution $P(v_{\perp},t)$. Dotted white lines show the cuts at time-values 
$t=0.548 T$, $t=0.7 T$, and $t=0.767 T$ for which the lateral velocity distributions shown 
in \Fref{f2} and \Fref{f3} were calculated.
}
\label{f1}
	\end{figure*}

We begin by presenting the results we obtain for the Yukawa potential
and field parameters $\omega=0.03$ a.u., $E_0=0.07$, which places 
us relatively deep in the tunneling regime. In the framework of the 
strategy we outlined above, we will analyze first the function 
defined in \Eref{ver} which will tell us at what particular
(if any) times inside the laser pulse the notion of the joint
probability can be sensibly used. This function is shown in
\Fref{f1}a. The plot shows a rather complicated
pattern, important for us are the "good" regions in the 
$(t,v_{\perp})$-plane, where $G(t,v_{\perp})$
is small. We need, of course, some quantitative criterion of "smallness". As such, 
we choose the condition $G(t,v_{\perp})\le 0.1$, which, as we will try to show, is sufficiently small threshold value 
for the notion of the joint probability distribution to make sense.  
One can see from \Fref{f1} that there are areas (shown in black) 
for times inside the laser pulse duration, where
this criterion is satisfied for the large enough intervals of lateral
velocities. 
In these areas, as we argued above, the joint probability
distribution is a sensible notion, and we can compute the 
lateral velocity distributions by taking cuts of the Cartesian
distribution $P(v_{\perp},t)$ we defined above in \Eref{vdist}
along the lines $t={\rm const}$. 

\begin{figure*}[!]
\resizebox{150mm}{!}{\epsffile{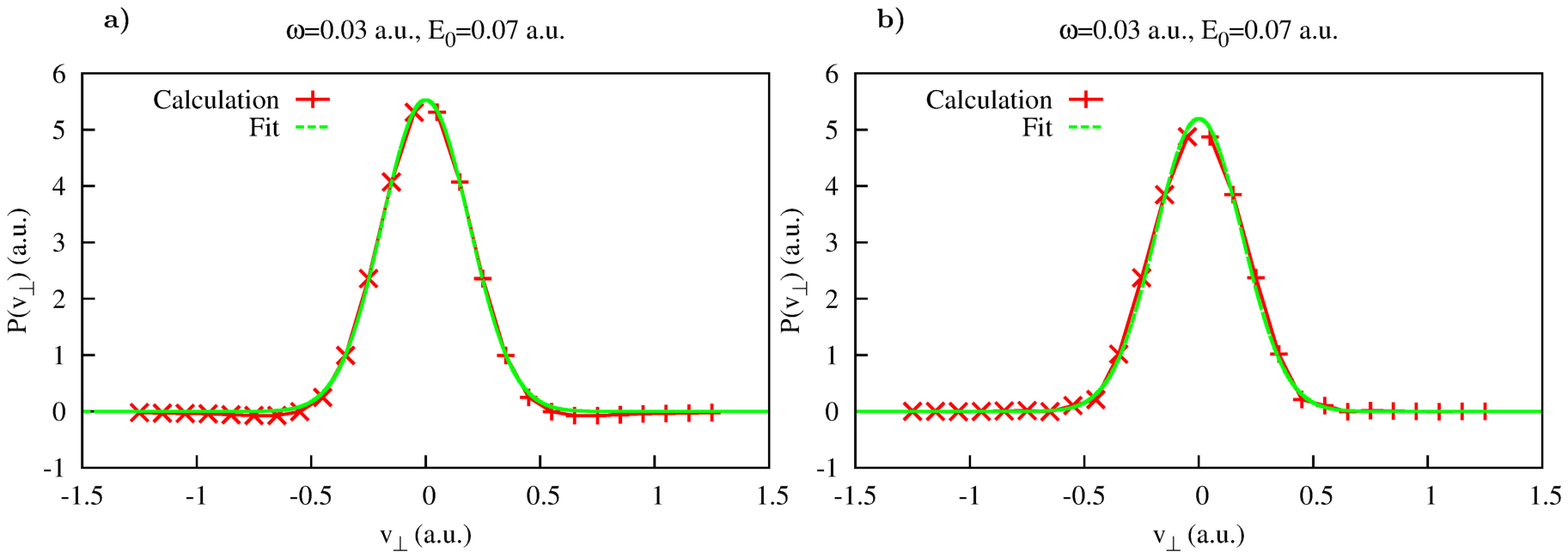}} 
\caption{(Color online) Yukawa potential. 
(a): Results of the fit of distributions computed according to \Eref{vdist} using \Eref{fitsr} as a fitting anzats. (a): $t=0.548T$. (b): $t=0.7T$. }
\label{f2}
	\end{figure*}

\begin{figure*}[!]
\resizebox{100mm}{!}{\epsffile{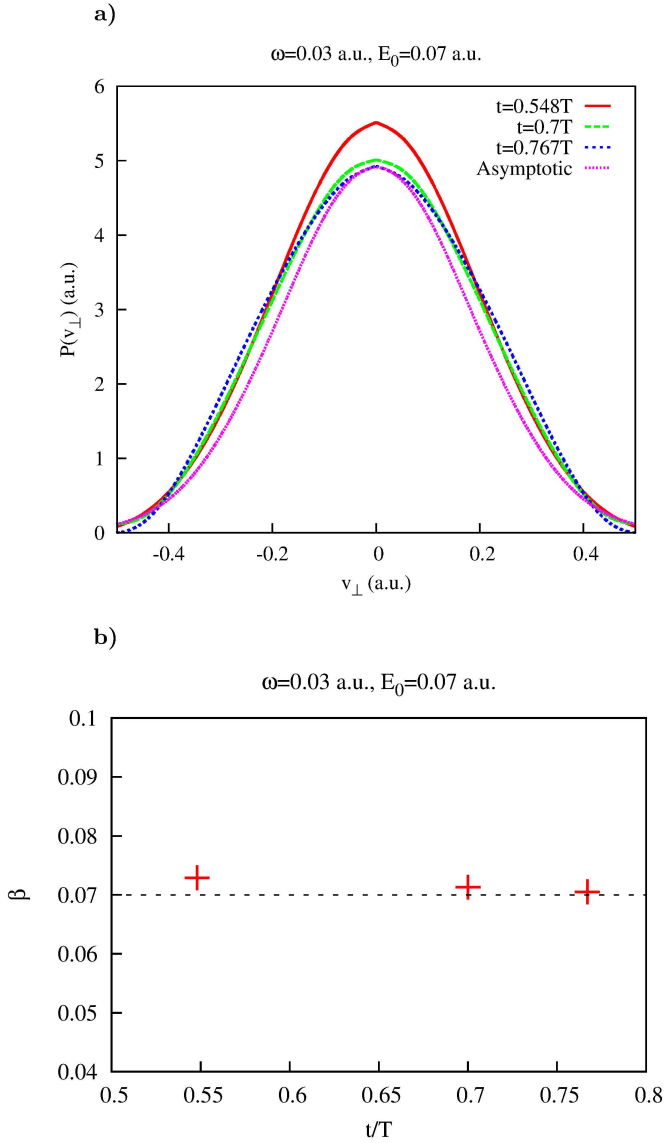}} 
\caption{(Color online) Yukawa potential. (a): 
Lateral distributions obtained using \Eref{vdist} for different moments of time inside the pulse duration (also shown as white dash lines in \Fref{f1}). Shown also is the asymptotic distribution obtained using \Eref{as}.  (b): Parameter $\beta$ of the fit as a function of time inside the interval of the laser pulse duration.}
\label{f3}
	\end{figure*}

The distribution $P(v_{\perp},t)$ is shown in \Fref{f1}b, and its cuts
$P(v_{\perp},t_i)$ taken along the lines $t=0.548 T$ and $t=0.7 T$
are shown in \Fref{f2}. As an inspection of the \Fref{f1} shows, for both these cuts
the areas where $G(t,v_{\perp})$ is small extend sufficiently far in $v_{\perp}$-direction
so that Cartesian lateral velocity distributions \eref{vdist} could 
be computed for the intervals of $v_{\perp}$ containing the bulk of the velocity 
distribution. The results shown in 
\Fref{f2} for the two moments of time inside the laser pulse duration show very nice agreement with the
Gaussian fit \eref{fitsr}  based on the SFA expression \eref{sfa}. The prediction given by the SFA
for the value of the parameter $\beta$ in \Eref{fitsr} is $\beta=0.07$ a.u.  
for the field strength $E_0$ we use. The values we obtain for $\beta$ by fitting the 
computed lateral velocity distributions are: $\beta=0.0729$ a.u. for $t=0.548T$ and 
$\beta=0.0710$ a.u. for $t=0.7T$. In \Fref{f3} we show evolution of the Cartesian lateral velocity
distributions for the three values of time inside the interval of the 
laser pulse duration (indicated by the dotted white lines in \Fref{f1}) and compare these distributions
to the asymptotic distribution $P^{as}(v_{\perp}) $ defined in \Eref{as}. As one can see from 
\Fref{f3}a, all these distributions, both the ones computed for times inside the interval of the 
pulse duration and the asymptotic distribution \eref{as} closely agree and shape of the 
distributions evolves little with time. This fact is illustrated also in \Fref{f3}b, where
we show evolution of the parameter $\beta$ in time. The pattern of the evolution we see
in \Fref{f3}b agrees completely with the picture provided by the SFA, where electron 
emerges into the continuum at the moment of the maximum field strength, with the lateral
velocity distribution given by \Eref{sfa}, which does not change subsequently. This latter
fact is due to the short range character of the potential which is assumed in the SFA,
and the fact that in the absence of any long range forces the electric field of the pulse
cannot affect the lateral components of electron's velocity.

\subsection*{Coulomb potential}

\begin{figure*}[!]
\resizebox{150mm}{!}{\epsffile{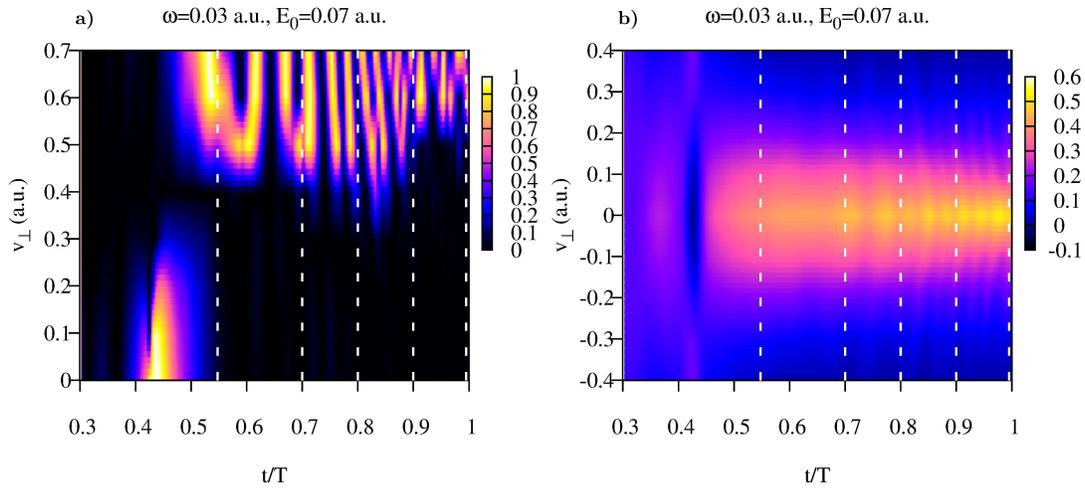}} 
\caption{(Color online) Coulomb potential. (a): Function $G(v_{\perp},t)$ 
(b): Distribution $P(v_{\perp},t)$. }
\label{f4}
	\end{figure*}

Results that we obtain for the Coulomb potential are somewhat different. This is to be expected as we have in
this case the strong long range Coulomb force. We use the same field parameters: 
$\omega=0.03$ a.u., $E_0=0.07$ as above. 
As in the case of the short range interaction, we proceed by calculating the function 
defined in \Eref{ver} to find the values of $t$ which could be used for
calculating lateral velocity distributions for the times 
inside the laser pulse duration.  
This function is shown in \Fref{f4}a.  For the Coulomb potential the region in the 
$(t,v_{\perp})$-plane where $G(t,v_{\perp})$ is small, 
proves larger than in the case of the Yukawa potential, 
allowing continuous scan of the velocity distribution for all times 
$t$ after the midpoint of the pulse.

\begin{figure*}[!]
\resizebox{150mm}{!}{\epsffile{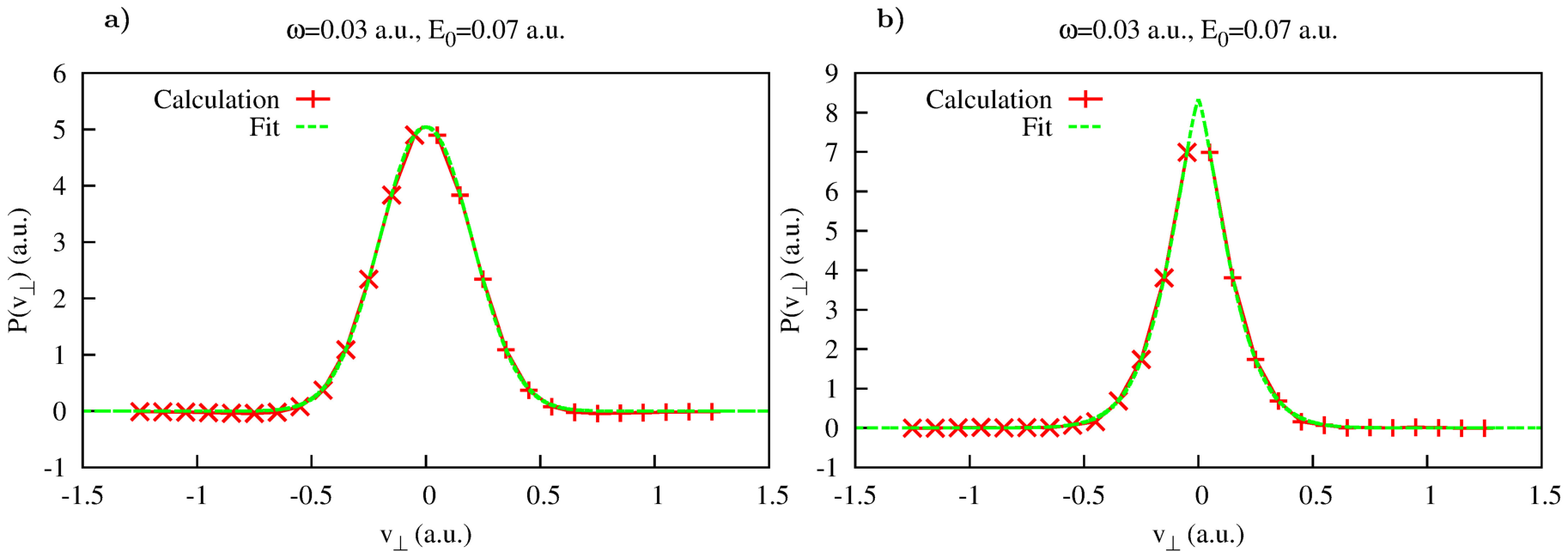}} 
\caption{(Color online) Coulomb potential. 
(a): Results of the fit of distributions computed according to \Eref{vdist} using \Eref{fitc} as a fitting anzats. (a): $t=0.548T$. (b): $t=T$. }
\label{f5}
	\end{figure*}

The distribution $P(v_{\perp},t)$ is shown in \Fref{f4}b, and the cuts
$P(v_{\perp},t_k)$ taken along the lines $t=0.548 T$ and $t=T$
are shown in \Fref{f5}. We also show results of the fitting procedure 
applied to the calculated distributions. The fitting procedure that we employ is
based on the expression \eref{fitc} which is non-analytical at
$v_{\perp}=0$, thereby taking into account presence of a cusp \cite{cusp2} for the Coulomb 
case \cite{cusp2}.

\begin{figure*}[!]
\resizebox{150mm}{!}{\epsffile{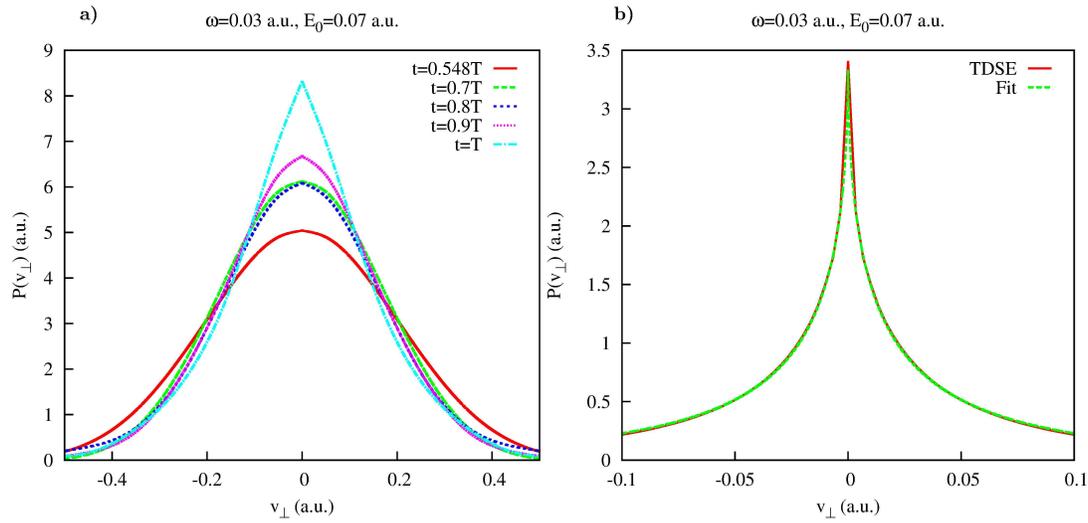}} 
\caption{(Color online) Coulomb potential. 
(a): Lateral distributions obtained using \Eref{vdist} for different moments of time inside the pulse duration (also shown as white dash lines in \Fref{f4}). (b): Asymptotic distribution obtained using \Eref{as}.}
\label{f6}
	\end{figure*}

As one can see, the fits based on \eref{fitc} reproduce fairly well the calculated 
lateral velocity distributions. One can also note that the distribution is more sharply peaked
at the end of the pulse than at the moment $t=0.548T$ near the midpoint of the pulse. 
This evolution of the lateral velocity distribution is presented in more detail in \Fref{f6}
for different moments of time inside the laser pulse duration. One can observe from \Fref{f6}a that
the non-analytical character of the velocity distribution and its departure from the Gaussian 
become progressively more pronounced as we approach the end of the laser pulse. This evolution of
the lateral velocity distribution does not stop there, of course, long range Coulomb force keeps 
distorting the velocity distribution long after the pulse is gone. The asymptotic lateral 
velocity distribution obtained using \Eref{as} is shown in \Fref{f6}b. One can see that it is
much sharper yet than the distribution we obtain for $T=1$. 
This evolution of the later velocity distributions we see in the Coulomb case 
agrees with the observation made in \cite{cuspo}, where it was noted that 
the cusp formation requires large (strictly speaking infinite) time. At any 
finite moment of time, the lateral velocity distribution remains an analytical function of 
$v_{\perp}$ which however, becomes, as time progresses, increasingly sharper 
in a vicinity of the point $v_{\perp}=0$ \cite{cuspo}, approaching 
thus the cusp-like behavior present in the asymptotic
(i.e. obtained for the infinite time) distribution shown in \Fref{f6}b.
This process of the cusp formation is illustrated in \Fref{f7}, where we show evolution of 
the fitting parameters in \Eref{fitc} with time. One can see that parameter $\alpha$ in 
\Eref{fitc} whose role in this expression is to mimic the cusp-like behavior, progressively 
decreases from the nearly Gaussian value $\alpha\approx 2$, at times close to the instant of ionization,
to the value $\alpha\approx 1.35$ at $t=T$.  Such a low value of $\alpha$ is responsible
for the rather sharp character of the lateral velocity distribution we observe for $t=T$ in \Fref{f6}a.
Indeed, the value $\alpha\approx 1.35$ at $t=T$ implies that the first derivative of 
the lateral velocity distribution at $v_{\perp}=0$ is continuous and finite, while
the second derivative is infinite. The asymptotic infinite time distribution shown in
\Fref{f6}b is sharper yet, we obtain the value $\alpha=0.55$ if we apply the fit based
on \Eref{fitc} to the distribution we calculate using \Eref{as}. That means that for the asymptotic 
distribution it is the first derivative which is infinite at $v_{\perp}=0$. These observations 
illustrate again the statement made in \cite{cuspo} to which we referred above, 
the cusp in the lateral velocity distribution becomes fully formed in the limit of infinite
time.

\subsection*{Field dependence of the lateral velocity distributions}

We present in this section the results we obtain for the lateral velocity distribution 
for different driving pulse parameters. The aim of performing 
these calculations was primarily to convince ourselves that the procedure 
we devised for the calculation of the lateral velocity distributions can be applied 
for a wide range of the laser field parameters. 
For the calculations presented below we use the frequency
$\omega=0.057$ a.u., and we vary the peak electric field strength. 
We follow the same strategy we used above and begin with presenting the analysis of the 
function $G(t,v_{\perp})$ defined in \Eref{ver}.

\begin{figure*}[!]
\resizebox{150mm}{!}{\epsffile{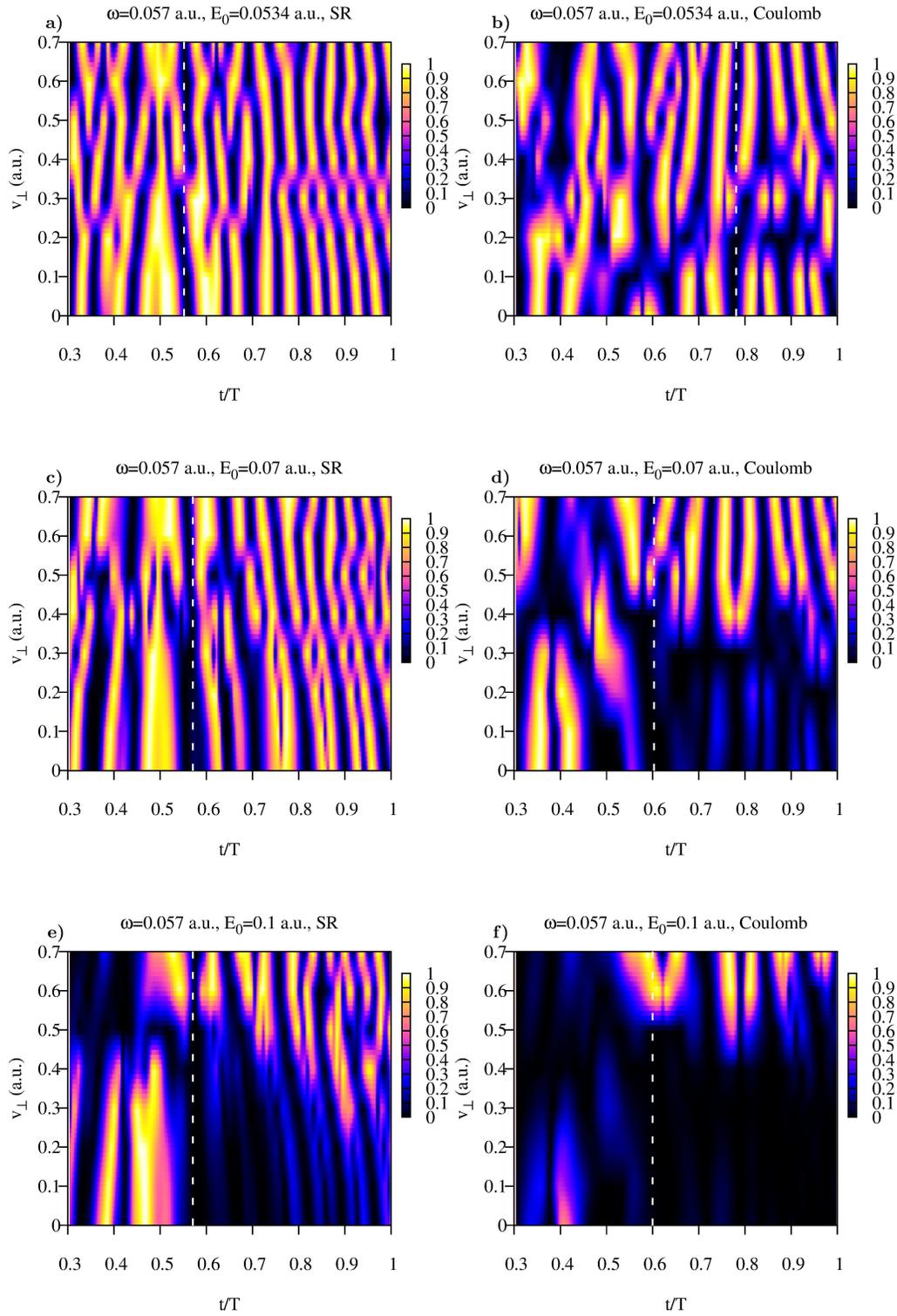}} 
\caption{(Color online) Coulomb and Yukawa potentials.  Function $G(v_{\perp},t)$ for different field strengths.}
\label{f8}
	\end{figure*}

This function is shown in \Fref{f8} for both Yukawa and Coulomb potentials. Choosing the 
values of $t$ where the function $G(t,v_{\perp})$ is small 
for the wide enough intervals of 
$v_{\perp}$ we can, as we did above, calculate the distributions 
defined in \Eref{vdist} for the range of the lateral velocities in 
which the distribution is predominantly concentrated.  

\begin{figure*}[!]
\resizebox{150mm}{!}{\epsffile{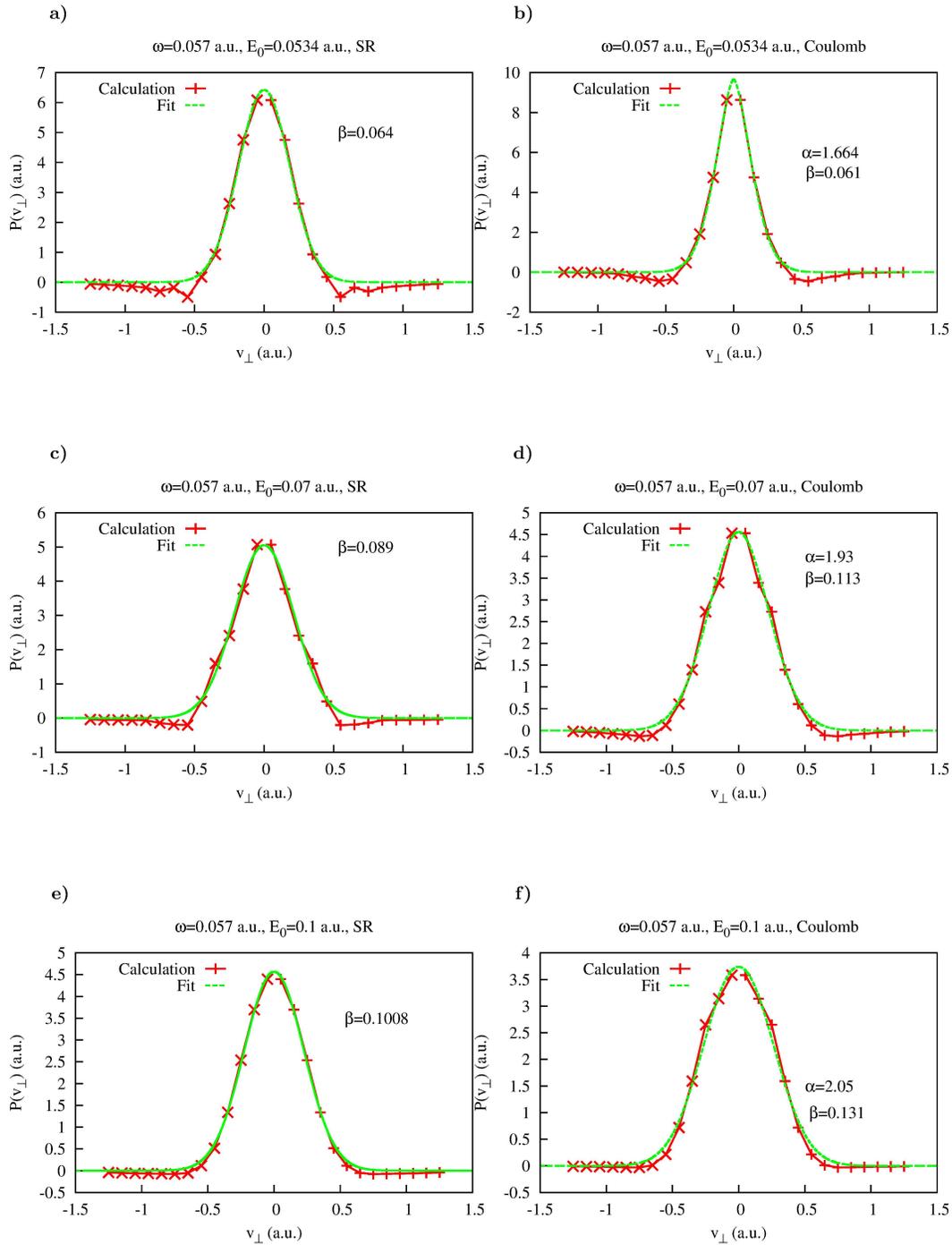}} 
\caption{(Color online) Coulomb and Yukawa potentials. 
Lateral distributions obtained using \Eref{vdist} for the following moments of time inside the pulse duration (also shown as white dash lines in \Fref{f8}). (a): $t=0.552T$, (b):$t=0.781T$, (c):$t=0.571T$, (d): $t=0.603T$,
(e):$t=0.571T$, (f): $t=0.609T$. }
\label{f9}
	\end{figure*}

The results of these calculations of the lateral velocity distributions  
for different driving field strengths 
are shown in \Fref{f9}. Also shown are the 
results of the fitting procedures based on \Eref{fitsr} and \Eref{fitc} for the 
Yukawa and Coulomb potentials, respectively. The fitting expressions
\eref{fitsr} and \eref{fitc} reproduce calculated distributions fairly well. 
The values of the fitting parameters we obtain are indicated in \Fref{f9}.
For the short range Yukawa potential the values of the fitting parameter $\beta$
are of primary interest. As one can see, these values indeed satisfy 
the relation $\beta\approx E_0$, which one would expect basing on the SFA formula 
\eref{sfa}. We see again that in the case of the 
short range interaction, the shapes and widths of the 
lateral distributions we obtain for the time moments
inside the laser pulse duration, coincide with the asymptotic distribution 
\eref{sfa}, describing velocity distribution at the detector. This 
agrees fully with the SFA-based scenario we mentioned above,  according to which the lateral 
velocity distribution does not evolve with time once electron is ionized because of the 
short range character of the atomic potential assumed in the standard form of the SFA.
For the Coulomb potential the fitting parameters are less informative. As we mentioned
above, the long range Coulomb forces exert considerable effect on the lateral velocity 
distributions so that the system must evolve for a long time for the 
distribution to approach its limiting asymptotic form.
We illustrate the character of this approach for the driving laser frequency $\omega=0.057$ a.u. in 
\Fref{f10}a, where we show time-evolution of the fitting parameter $\beta$ in \Eref{fitc}.
The asymptotic distribution \eref{as} and its fit \eref{fitc} are shown in 
\Fref{f10}b. One can see that after the ionization event $\beta$ starts decreasing,
so that the lateral velocity distribution becomes progressively sharper at
$v_{\perp}=0$.  At the end of the pulse $\beta\approx 1$, which is still far from the 
value $\beta\approx 0.59$ which our fitting procedure \eref{fitc} gives for 
the asymptotic distribution shown in \Fref{f10}b. We see thus the same behavior of the 
cusp formation that we saw in \Fref{f7} for the driving laser frequency $\omega=0.03$ a.u. The cusp 
is not fully formed at the end of the pulse, the parameter characterizing the sharpness
of the cusp needs more time yet to reach its asymptotic value. 

\section*{Conclusion}

We presented a study of the lateral velocity distribution and its development in time
during the interval of the laser pulse duration. Our approach is based on the application of
the notion of the joint probability. Though this notion is not very well defined
in QM and may sometimes lead to unphysical (e.g. complex) probability distributions,
one may find situations when such a joint probability becomes a meaningful concept.
This opens a way to tackle the questions which are somewhat difficult to
address, in particular, to study the characteristics of the ionized electron for times
inside the interval of the laser pulse duration. As we mentioned above, 
for the moments of time inside the interval of the laser pulse duration, when 
the wave-packet describing ionized electron is not fully formed and is still partly inside 
the atom, it is not easy to unambiguously single out the part of the wave-function describing ionized
electron from the total wave-function of the system. 

In the framework of our approach this ambiguity is resolved by calculating joint probability of
two events, event $A$ being detection of the lateral electron's velocity in a given volume 
of the electron's velocity space and event $B$ being electron's detection by a detector placed far away 
from the atom. We saw, that though the joint probability thus defined
cannot be used for all moments of time inside the laser pulse duration, since operators
corresponding to the observables $A$ and $B$ generally do not commute, 
one can find the instances when these operators commute approximately, allowing us to define 
sensible joint probability distribution of the events $A$ and $B$.

\begin{figure*}[!]
\resizebox{100mm}{!}{\epsffile{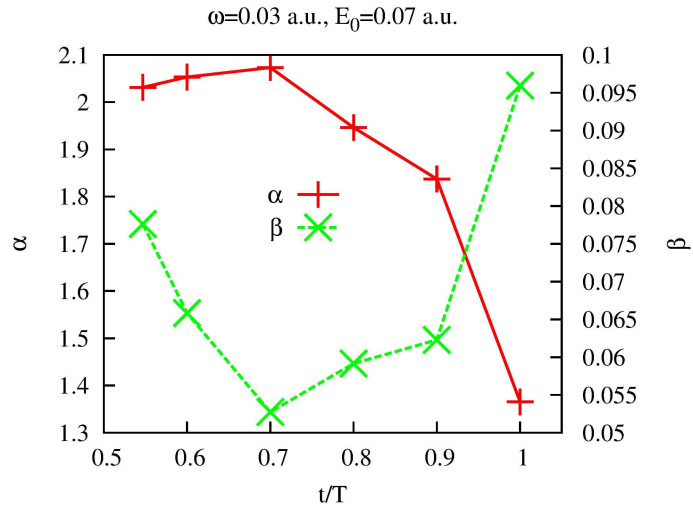}} 
\caption{(Color online) Coulomb potential. Parameters
$\alpha$ and $\beta$ of the fit based on \Eref{fitc} as functions of time inside the interval of the laser pulse duration.}
\label{f7}
	\end{figure*}

\begin{figure*}[!]
\resizebox{150mm}{!}{\epsffile{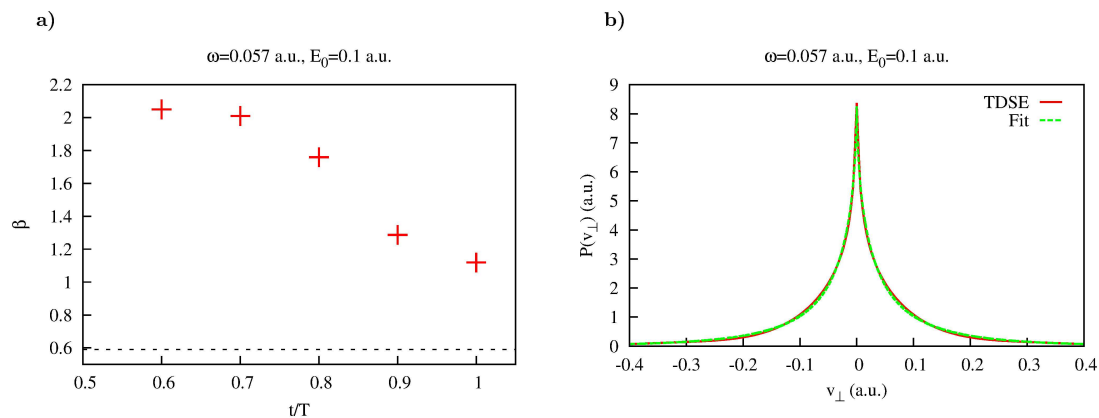}} 
\caption{(Color online) Coulomb potential. (a): Parameter
$\beta$ of the fit based on \Eref{fitc} as function of time inside the interval of the laser pulse duration. (b): Asymptotic lateral velocity distribution obtained using \Eref{as}. }
\label{f10}
	\end{figure*}

By following this strategy we were able to track the time-evolution of the lateral velocity 
distributions on the time interval between the moment of ionization and the end of the laser
pulse. We saw that in the case of the short range Yukawa potential, this evolution 
reproduces the standard SFA scenario. The shape of the lateral velocity 
distributions for times inside the laser pulse is reproduced pretty accurately by the 
SFA formula \eref{sfa}, and it changes very little after the moment of ionization. This
is, of course, a consequence of the short range character of the Yukawa potential, which 
has little influence of the motion of the ionized electron while electric field alone cannot
alter velocities in lateral directions. We observe quite different behavior in the case of 
the Coulomb potential, which is expected since long range Coulomb force can, strictly speaking,
never be neglected. Since the average of the electric field 
over an optical cycle is zero, the long range Coulomb force may affect electron's motion 
to a considerable degree even when electron is far away from the ionic core. 
This effects manifests itself, in particular, in the process of the time-development of the cusp in
the lateral velocity distributions in the Coulomb case.
We saw that the lateral velocity distributions we compute 
using \Eref{vdist} become progressively more sharply
peaked near the point $v_{\perp}=0$, as illustrated in \Fref{f7} and \Fref{f10}, thus
showing the development of a cusp. Nevertheless, even at the end of the pulse this
development is far from being complete, the values of the parameter $\beta$ characterizing
"sharpness" of the cusp are still far from the value we retrieve from the asymptotic
(infinite time) distribution \eref{as}. 

In the present work we used as objects of study 
the simple one-electron targets, the hydrogen atom and its counterpart described by the 
short-range Yukawa potential with the same ionization potential, in the field of a linearly polarized electromagnetic 
wave. The theoretical framework we used can be applied for more complicated targets as well, such as 
heavier atomic species or multiply charge ions with larger ionization potential. 
We would expect qualitatively similar results
for these targets. In the absence of the long range Coulomb force the formation of the lateral velocity 
distributions, as we saw, is very well captured by the SFA, which does not rely on any information about
atomic target at all, except the value of the ionization potential. We believe, therefore, that 
for the targets with higher ionization potential, described by the short range forces, we should
obtain results similar to the results we presented above for the Yukawa SR potential, with 
the shape of the lateral velocity distribution reproduced fairly well 
for times inside the laser pulse by the 
SFA formula \eref{sfa} and changing insignificantly after the moment of ionization. 
For the targets with the ionic potential exhibiting long range Coulomb tail (e.g. the 
multicharged ions) we would expect stronger yet effect of the Coulomb field on the cusp formation.
That should give us the overall picture of the cusp formation qualitatively similar to the 
case of hydrogen we studied above, resulting in a considerable evolution of the cusp parameter in
the lateral velocity distribution, like the one shown in \Fref{f10}. For the 
systems with stronger Coulomb forces the 
particular values of the parameters characterizing the cusp can, of course, be completely
different from those shown in \Fref{f10}. 

We can apply this technique to other field geometries
as well, in particular to the case of the circularly or elliptically polarized electromagnetic 
fields. The case of the driving field polarization different from the linear one is 
of particular interest since for elliptically polarized field
the tunneling electronic wave packet possesses an initial transverse
momentum due to the non-adiabatic effects \cite{tol1,tol2}. Calculations for such field
geometry are considerably
more time consuming because of the necessity of performing multiple solutions 
of the TDSE for the much more computationally 
demanding case of the non-linear polarization. We plan, nevertheless, to perform such calculations in
the future to study these highly interesting non-adiabatic effects.

\begin{figure*}[!]
\resizebox{90mm}{!}{\epsffile{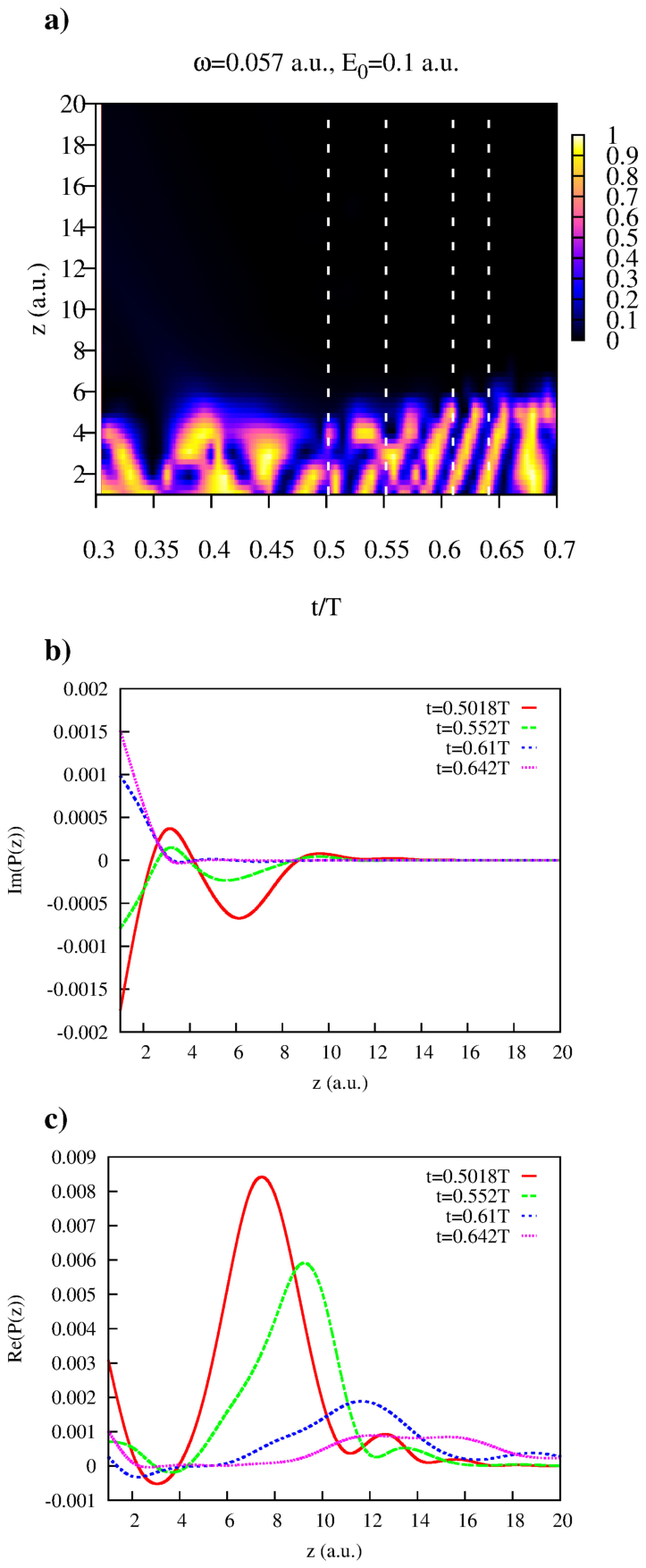}} 
\caption{(Color online) Yukawa potential. (a): Function $G(z,t)$.
Dotted white lines show the cuts at time-values 
$t=0.5018T$, $t=0.552T$, $t=0.61T$, and $t=0.642T$ for which the imaginary (b), and 
real (c) parts of the joint distributions $P(z)$ 
were calculated.}
\label{f11}
	\end{figure*}

We also note that our approach is by no means limited to the study of the lateral 
velocity distributions only. By choosing other observables to represent event $A$ at
times inside the interval of the laser pulse duration and applying the strategy we described above,
we can study evolution of the corresponding distributions. As an illustration of this statement we
consider briefly some results we obtain for another choice of the event $A$. 
Let us define $A$ as an event consisting in finding the electron in a
region around the point $\r_0$ in space at the moment of time $t$. 
More specifically, we define the projection operator 
$\hat Q_{\r_0}$ representing this observable as: 
$\hat Q_{\r_0} = |\phi_{\r_0}(\r)\rangle \langle \phi_{\r_0}(\r)|$.  For $|\phi_{\r_0}(\r)\rangle$ 
we use the Gaussian form: $\displaystyle \phi_{\r_0}(\r) = N e^{-a(\r-\r_0)^2)}$, where 
$N$ is the normalization factor and we use the value $a=2\ln2$ for the parameter $a$. This parameter 
defines the 'resolution' with which we can scrutinize the coordinate space, and it is approximately 
one atomic unit of length for the choice of the parameter $a$ we made. In the \Eref{c4} 
we can replace now the projection operator 
$\hat Q_i$ with the projection operator $\hat Q_{\r_0}$, and the resulting expression
will give the joint probability $P(\r_0)$ of finding electron ionized after the end of the pulse and 
finding it in a location nearby $\r_0$ at the moment $t$. In other words, the joint 
probability thus calculated gives us a distribution of the ionized electron coordinates at
times inside the laser pulse. We proceed with the calculations in exactly the same way as we did before,
obtaining the results shown in \Fref{f11} for the case of the short range Yukawa potential. 
We use the same pulse parameters as above and report the results for the peak field strengths $E_0=0.1$ a.u.
To simplify the notation we drop the subscripts
and we will write simply $\r$ instead of $\r_0$ in the formulas below which should not cause confusion.

We show in \Fref{f11}a the 
function $G(z,t)$ calculated according to \Eref{ver}, with projection operator $\hat Q_k$ replaced with 
the operator $\hat Q_{\r}$ for $\r=(0,0,z)$. 
We remind that in our geometry $z$-direction is the polarization direction, so by choosing this direction
in space we can study the coordinate distribution for the ionized electron along the laser polarization 
direction. One can see from \Fref{f11}a that for sufficiently large $z$ and all $t$ $G(z,t)$ is small.
Joint distributions computed according to the \Eref{c4} give us, therefore, reliable 
distributions for the ionized electron $z-$ coordinate for all $t$ and at least some interval of 
$z$-values. This can also be seen from \Fref{f11}b and \Fref{f11}c where we show imaginary and 
real parts of the joint distributions $P(z)$ computed according to \Eref{c4}. One can see that
for the values of $z$ of interest (near the peaks of the real parts of $P(z)$) imaginary part
of $P(z)$ is at least an order of magnitude smaller than the real part. We obtain thus
the meaningful ionized electron's coordinate distributions for the most important and 
interesting intervals of $z$, where the bulk of a distribution is concentrated. As one can
see from \Fref{f11}c electron ionized at the moment near the field maximum at $t=0.5T$ is located 
at $z\approx 7$ a.u. This is not very different from a simple estimate for the tunnel exit location
based on the energy conservation 
formula for an electron moving in the combined field of the ionic core and 
the static electric field with the amplitude $E_0$ (the 
so-called Field Direction Model (FDM) \cite{landsman2015}), which, for the Yukawa potential we consider
and $E_0=0.1$ a.u. is $z\approx 5$ a.u. 

Using this approach one 
can obtain information about other nonlinear phenomena as well. The rescattering process, for instance,  
is at the core of the 
HHG process and is responsible for the formation of the high energy part of the ATI spectra \cite{kri}.
Consider, for instance, the ATI process. By choosing
event $B$ to be detection of the electron in the ionized state and 
event $A$ to be finding the electron's coordinate in a certain region of space (as we did above
studying the ionized electron's coordinate distributions), 
we can single out the part of 
the wave-function describing the ionized wave-packet, and study
rescattering process in detail by following spatial and temporal evolution of the 
ionized wave-packet.

\section*{Data availability}
All relevant data are available from the authors upon request sent to I.A.I.

\section*{Acknowledgements}
 
This work was supported by IBS (Institute for Basic Science) under IBS-R012-D1.

\section*{Author contributions statement}
I.A.I. and  K.T.K conceived the project. I.A.I. performed the calculations. 
All authors contributed to discussions.

\section*{Additional information}

\textbf{Competing interests:} 
The Authors declare no Competing Financial or Non-Financial Interests.


\end{document}